\documentclass[prl,twocolumn,aps,amsmath,amssymb,showpacs,superscriptaddress]{revtex4}%

\usepackage{graphics}
\usepackage{dcolumn}
\usepackage{bm}
\usepackage[T1]{fontenc}

\begin{document}

\title{Shot Noise in Ballistic Graphene}

\author{R. Danneau}\email[Corresponding author: r.danneau@boojum.hut.fi]{}
\author{F. Wu}
\affiliation{Low Temperature Laboratory, Helsinki University of Technology, Espoo, Finland}
\author{M.F. Craciun}
\author{S. Russo}
\affiliation{Kavli Institute of Nanoscience, Delft University of Technology, Delft, The Netherlands}
\author{M.Y. Tomi}
\author{J. Salmilehto}
\affiliation{Low Temperature Laboratory, Helsinki University of Technology, Espoo, Finland}
\author{A.F. Morpurgo}
\affiliation{Kavli Institute of Nanoscience, Delft University of Technology, Delft, The Netherlands}
\author{P.J. Hakonen}
\affiliation{Low Temperature Laboratory, Helsinki University of Technology, Espoo, Finland}

\begin{abstract}

We have investigated shot noise in graphene field effect devices in
the temperature range of 4.2--30 K at low frequency ($f$ = 600--850 MHz). We find that for our graphene samples with large width
over length ratio $W/L$, the Fano factor $\mathfrak{F}$ reaches a
maximum $\mathfrak{F} \sim$ 1/3 at the Dirac point and that it
decreases strongly with increasing charge density. For smaller
$W/L$, the Fano factor at Dirac point is significantly lower. Our
results are in good agreement with the theory describing that
transport at the Dirac point in clean graphene arises from
evanescent electronic states.

\end{abstract}

\pacs{}

\maketitle

Recently, graphene, a two-dimensional carbon crystal arranged in
honeycomb lattice, has attracted tremendous attention in the
scientific community \cite{geim2007}. Its fundamental physical
properties open up possibilities for realizing high-speed, ballistic
field effect transistors \cite{miao2007} as well as for potential
applications like spin control devices \cite{trombos2007} or gas
sensors \cite{schedin2007}. Owing to its unique band structure, low
energy conduction in graphene occurs via massless Dirac fermion
quasiparticles leading to very rich and sometimes counterintuitive
behavior \cite{geim2007}.
For example, despite the vanishing density of states at the Dirac
point where the conduction and the valence band touch, the
conductivity of graphene remains finite. Indeed, it has been proposed that
transport at the Dirac point can be explained as propagation of charge
carriers via evanescent waves \cite{tworzydlo2006}. Universal minimum
conductivity of $\frac{4e^{2}}{\pi h}$ has been observed around the
Dirac point for samples with large width over length ratio ($W/L$)
\cite{miao2007}, yielding support to this evanescent-wave theory
\cite{tworzydlo2006}. In addition to the minimum conductivity,
current fluctuations are expected to exhibit non-trivial
characteristics. A universal maximum value of 1/3 for the Fano
factor $\mathfrak{F}$, decreasing with increasing charge carrier
density, is expected for large $W/L$ in graphene. However, no
experimental results have been reported yet \cite{dicarlo2007}. In this Letter,
we present a study of shot noise in short graphene strips at
frequencies $f= 600 - 850$ MHz, which is well above the $1/f$ corner
frequency, even at the large currents when working far away from the
Dirac point \cite{chen2007}. Using a high
resolution noise measurement set-up \cite{wu2006}, we measure shot
noise of two-terminal graphene devices both as a function of the
charge density as well as bias voltage. Our results indicate that
ballistic transport occurs at the Dirac point via evanescent waves.

\begin{figure}[htbp]
\scalebox{0.34}{\includegraphics{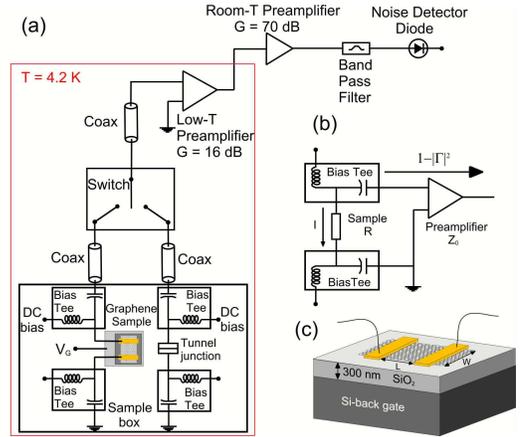}}\caption{(a)
Experimental set-up for detecting shot noise at $T$ = 4.2--30 K.
(b) Schematic of the principle of our measurements in terms of the
noise power reflection $|\Gamma|^2$. (c) Illustration of a typical
graphene sample fabricated for our shot noise study.}
\end{figure}

The role of disorder, interactions or carrier statistics in quantum
transport can be probed by shot noise in mesoscopic devices
\cite{blanter2000}. Arising from the granular nature of electron
charges, shot noise reveals information on fundamental conduction
properties of low-dimensional systems in a complementary manner to
conventional dc transport measurements. Shot noise can be detected only
when the drift time is shorter than electron-phonon energy relaxation time
\cite{beenakker1992,steinbach1996}. Naively, since electrical transport
in graphene at submicron scale is believed to be ballistic, one
would expect shot noise to be completely absent, \emph{i.e.} the Fano factor
$\mathfrak{F} = \sum_{n=1}^{N}T_n(1-T_n)/\sum_{n=1}^{N} T_n$ = 0, where $T_n$
are the transmission coefficients of the conducting channels in the scattering matrix formalism \cite{blanter2000}
(when $\mathfrak{F}$ = 0, all eigen values of the transmission matrix are equal to 1).
However, Tworzyd{\l}o \emph{et al.}
\cite{tworzydlo2006} have calculated that perfect graphene,
contacted by highly doped graphene leads, should
exhibit a Fano factor of 1/3 at the Dirac point for high $W/L$ (note that Katsnelson reported similar sub-poissonian shot noise in \cite{katsnelson2006a}).
Similar calculations using metallic square lattice
contact lead to the same conclusions \cite{schomerus2007}. In Ref.
\onlinecite{tworzydlo2006}, the authors show that the transmission
coefficients $T_n = \cosh^{-2}[\pi n\frac{L}{W}]$, \emph{i.e.} they
display the same distribution as in coherent disordered conductors.
No clear-cut explanation for this intriguing coincidence between
transmission eigenvalue distribution for the propagating modes in
diffusive systems and evanescent states in graphene has been given
\cite{tworzydlo2006}.

Our experimental setup is illustrated in Fig. 1(a). In our work, we use
the sensitive lock-in detection technique described in Refs.
\onlinecite{wu2006,wu2007}. To enhance the sensitivity of noise
detection, we use a low-frequency sine-wave modulation of the
current, $I = I_{dc} + \delta I \sin(\omega t)$ where $I_{dc}
\gg \delta I$.
In order to avoid external spurious signals, the set-up is placed in
a Faraday cage and the signal is band limited to the range of 600--850 MHz,
thus avoiding EMI from mobile phone frequencies. We measure
the shot noise of a tunnel junction ($\mathfrak{F}$ = 1) for
calibration. A microwave switch is used to alternatively measure the
noise from the graphene sample and the tunnel junction. Bias-tees
are used to split dc bias and the bias-dependent high-frequency
noise signal. The noise signal is first amplified by a low-noise
amplifier (LNA) with a noise temperature of $T_n$ = 3.5 K in matching conditions,
thermalized at the same temperature as the sample,
then by a series of room-temperature amplifiers,
and finally collected by a zero-bias Schottky diode with band-pass
filtering of $f$ = 600--850 MHz. The noise power measured at the
output of the LNA is a mixture of thermal noise of the LNA and shot
noise from the sample. It can be defined in terms of the reflected
signal $|\Gamma|$ (see Fig. 1(b)). Here $\Gamma =
\frac{R-Z_{0}}{R+Z_{0}}$ is the signal reflection coefficient
when the noise generator, \emph{i.e.} the measured sample with a
resistance $R$, does not match to cold amplifier with the impedance
$Z_{0}$ = 50 $\Omega$. Then, the measured noise power reads:
$P(I) = P_{noise}(1-|\Gamma|^{2}) = \mathfrak{F} \times
2eI \times 4Z_0\left(\frac{R}{R+Z_0}\right)^{2}$ ,
%\begin{eqnarray}
%$P(I) &=& P_{noise}(1-|\Gamma|^{2}) = \mathfrak{F} \times
%2eV\frac{4RZ_0}{(R+Z_0)^{2}} \nonumber \\ &=& \mathfrak{F} \times
%2eI \times 4Z_0\left(\frac{R}{R+Z_0}\right)^{2}$ ,
%\end{eqnarray}
where $P_{noise}=\mathfrak{F}2eIR$ is the shot noise generated by the sample at $T=0$.
When $R \gg Z_0$, the coupling factor,
$\left(\frac{R}{R+Z_0}\right)^{2}$ can be taken as $=1$. In our
graphene samples, this is not fully true, and we employ the
differential resistance measured at bias voltage $V_{bias}$ for
calculating the shot noise of the source from the measured noise
power.
The lock-in detection yields the noise derivative
$P_d=\frac{dP}{dI}$, which can be integrated to
yield the average Fano factor $F=\frac{1}{P_{cal}}( \int_0^{I} P_d dI)/(2eI)$ where
 $P_{cal}$ denotes the separately determined calibration constant. Thus,
 in terms of current noise, our average Fano factor corresponds
 to $F=(S_I (I)-S_I (0))/(2eI)$ \cite{wu2007}.

The schematic of a typical sample is shown in Fig. 1(c). Graphene
sheets were mechanically exfoliated
using the Scotch-tape technique \cite{geim2007} and transferred from the natural graphite
crystals to the surface
of SiO$_{2}$/Si substrate (300 nm thick thermally grown SiO$_{2}$
layer). The heavily doped Si substrate is used as a back-gate in the measurements.
Single graphene layers were
located using a 3CCD camera in an optical microscope on
the basis of the RGB green component shift, in a similar manner as
was recently done in Ref. \onlinecite{oostinga2008}. After standard
e-beam lithography, a bilayer of Ti (10 nm) / Au (40 nm) was
evaporated, followed by lift-off. Ti was chosen because it allows the formation
of highly transparent contacts to graphene. In fact, in previous
work on graphene superconducting junctions with Al/Ti contacts \cite{heersche2007}, a very large
probability for Andreev reflection was observed indicating an average
transmission probability at the Ti/graphene interface $\geq$ 0.8. This is important because in our two lead configuration, we are sensitive to
total transmission eigenvalues of both graphene and graphene/leads interface.

\begin{figure}[htbp]
\scalebox{0.25}{\includegraphics{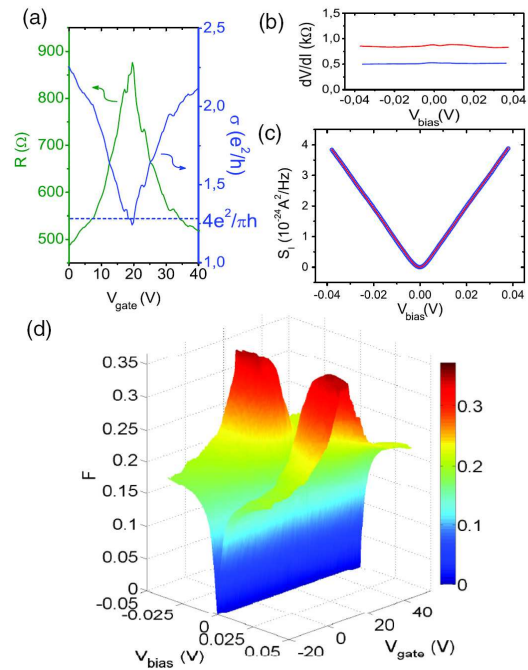}} {\caption{Measurements on
sample with $W/L$ = 24: (a)  Resistance $R$ (left axis) and
conductivity $\sigma$ (right axis) as a function of $V_{gate}$. (b) Differential resistance $dV/dI$ versus bias voltage $V_{bias}$ at the Dirac point (red curve) and at high density (blue curve). (c) Current noise per unit bandwidth $S_{I}$ as a function
of bias at the Dirac point, at $T$ = 8.5 K, fitted (red curve) using Khlus formula ($\mathfrak{F}
= $ 0.318). Note that the low bias data are perfectly fitted as
well as the high bias (d) Mapping of the average Fano factor $F$ as a function of gate voltage $V_{gate}$ and bias voltage $V_{bias}$ at $T =$
8.5 K.}}
\end{figure}

All the data was measured in a helium dewar, in which samples
were in a He-gas atmosphere of 1 bar. The resistance of the samples was measured using standard
low-frequency ac lock-in technique with an excitation amplitude of 0.3
mV ($\sim$ 3 K) at $\omega/2\pi = 63.5$ Hz, in the temperature
range of 4.2--30 K. As shown in Fig. 2(a), the sample resistance
as a function of the charge carrier density shows a maximum at the
Dirac point for our large aspect ratio sample ($W/L =
24$ with $L \sim 200$ nm). Our graphene samples are
non-intentionally \emph{p}-doped, probably due to oxygen gas adsorption \cite{schedin2007}. As a consequence, the
Dirac point is shifted to a positive $V_{gate}$ value.
Nevertheless, from our  measured conductivity
values, it seems that adsorbed gas on a graphene sheet does not
affect dramatically the transport properties of our samples. For our $W/L =
24$ sample, we obtain a minimum conductivity around $\frac{4e^{2}}{\pi h}$
which is the one calculated for large aspect ratio
\cite{tworzydlo2006} and measured in recent experiments
\cite{miao2007}. Note that the resistance of
graphene is nearly independent of the bias voltage $V_{bias}$ (see Fig. 2(b)).

Fig. 2(c) displays the power spectral density of current noise as a
function of the $V_{bias}$ at $T$ = 8.5 K for our $W/L = 24$
sample, at the Dirac point. The first way to extract the Fano
factor $\mathfrak{F}$ is to use the formula originally introduced by Khlus
\cite{khlus1987} which describes the cross-over from thermal to shot noise when $eV
\sim k_{B}T$ \cite{note1},
%\begin{eqnarray}
$S_{I}=\mathfrak{F}\frac{2e|V|R}{\left(R+Z_{0}\right)^2}
\coth\left(\frac{e|V|}{2k_{B}T} \right)\label{equa3}$ . %\frac{4k_{B}T_{noise}}{Z_{0}}+
%\end{eqnarray}
Here $Z_{0}$ is the impedance of the cold LNA (50 $\Omega$). Since the resistance of our graphene samples is bias
independent, we may fit the Khlus formula to our data using only
$\mathfrak{F}$ as a fitting parameter at fixed temperature $T$. From our nearly perfect fit,
we extract that $\mathfrak{F} = 0.318$ at $T = 9$ K . To check the
accuracy of our measurements, we also used the tunnel junction
calibration technique to extract the average Fano factor $F$
\cite{wu2006,wu2007}. We found $F$ = 0.338 at the Dirac point (at $V_{bias}$ = 40 mV).

The values found for $\mathfrak{F}$ and $F$ as well as for the minimum conductivity
are very close to the expected theoretical value of 1/3 and $\frac{4e^{2}}{\pi h}$ respectively
at the Dirac point for perfect graphene strip with large $W/L$ \cite{tworzydlo2006}.
However, the Fano factor in this case is also the one expected for a diffusive mesoscopic system.
In order to confirm that charge carriers in our sample do not undergo any scattering,
we swept $V_{gate}$ to move the Fermi level across the Dirac point to high carrier density.
A diffusive system should not display any gate dependence (as measured by DiCarlo et al. in \cite{dicarlo2007}
and calculated for long range disorder in \cite{lewenkopf2007}).
In Fig. 2(d), we can see a mapping of $F$,
calculated by integrating the differential Fano factor as described
in \cite{wu2007}, as a function of the $V_{bias}$ and $V_{gate}$. It
shows a clear dependence of $F$ on gate voltage \cite{note2}(i.e. the charge carrier density), with a clear drop (about a factor of 2) of the Fano factor at large carrier density
confirming that our results are in good agreement with the evanescent state theory \cite{tworzydlo2006}.

It is important to note that the Fano factor is barely affected by
temperature indicating that both the length $L$ and the width $W$
are smaller than the electron-phonon inelastic scattering length
$L_{e-ph}$: if this condition was not fulfilled, the Fano factor
would decrease, approximatively as inversely proportional to $N =
\frac{L_{e-ph}}{ \max (W,L)}$, though the actual form
would be model dependent \cite{blanter2000}. Since our shot noise measurements do not depend on temperature (between 4 and 30 K)
and with our contacts being highly transparent, the presence of
inelastic scattering mechanism in the graphene sample and at its interfaces with the leads can be ruled out. Note that bad contacts can only increase the Fano factor toward the limit of two symmetrical tunneling barriers in series which gives a Fano factor of 1/2. This is not the case of our samples in which the Fano factor has never been measured higher than 1/3.

\begin{figure}[htbp]
\scalebox{0.19}{\includegraphics{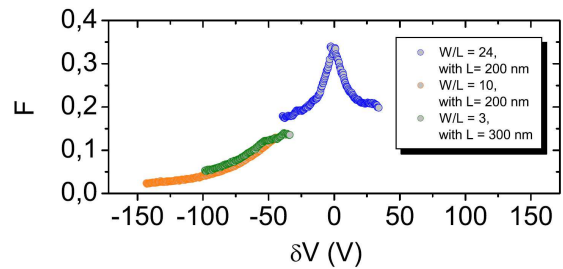}}{\caption{$F$ extracted at $V_{bias}$ = 40 mV for three
different samples, all having $W/L \geq$ 3, as a function of $\delta V = V_{gate} - V_{Dirac}$. For
the two unintentionally highly \emph{p}-doped samples (orange and green
dots), the Dirac point was estimated via extrapolation of the
minimum conductivity at $\frac{4e^2}{\pi h}$.}}
\end{figure}

In Fig. 3, we have plotted the $F$ as a function of
$\delta V = V_{gate} - V_{Dirac}$
for three different samples with $W/L\geq$ 3 (3, 10 and 24).
All samples were \emph{p}-doped, the Dirac points being at positive
$V_{gate}$, but only for one of these three samples we could reach
the Dirac point (sample with $W/L = 24$). The gate voltages
corresponding to the Dirac point for the two other samples were
estimated from their conductivity curves. We observe that
despite the high doping level of the samples, the Fano factor seems
to behave universally and tends to zero at very high density,
i.e. the value for a ballistic system.

We found that the value for the Fano factor for high
$W/L$ is about 1/3 to a good accuracy ($\pm$ 5 \%),
indicating that transport occurs via evanescent modes without any
observable resonant scattering at the Dirac point.
It has been theoretically demonstrated that the
presence of disorder-induced scattering in graphene decreases the Fano
factor while  disorder, counterintuitively, enhances
conductivity \cite{titov2007,sanjose2007}. Such behavior may be
understood as consequences of the absence of intervalley scattering
\cite{morpurgo2006} and the chirality conservation
\cite{katsnelson2006b}. Recently, a  model calculation near
the Dirac point was performed  \cite{sanjose2007} for conductivity
and shot  noise taking into account smooth disorder, namely
puddles \cite{hwang2007,martin2007} generated by charged impurities.
According to this model, smooth disorder at length scales $\ll L,W$ lowers the Fano
factor at the Dirac point, down to 0.243 for one-dimensional disorder and to
0.295 for the two-dimensional case \cite{sanjose2007}. This value falls clearly below
our experimental result $\mathfrak{F}$ = 0.33 $\pm$ 0.02   indicating
that transport occurs via evanescent modes without any
observable scattering at the Dirac point.
Despite the probable presence of some disorder in our system,
the transport regime can be considered as ballistic on our sample length scale.
We note that weak disorder may induce
anomalously large conductance fluctuations \cite{rycerz2007}.
However, we do not observe any oscillations in the conductivity or
in the Fano factor by increasing density.

\begin{figure}[htbp]
\scalebox{0.6}{\includegraphics{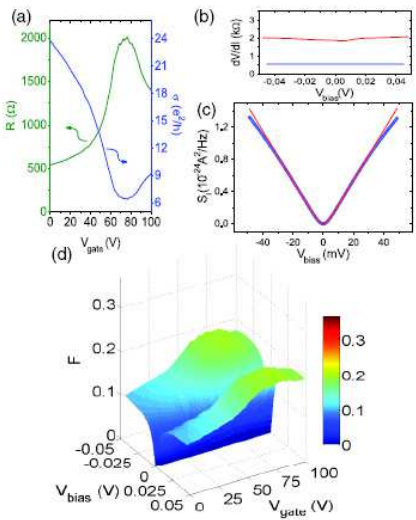}} {\caption{Measurements
on sample with $W/L$ = 2: (a) $R$ (left axis) and
$\sigma$ (right axis) as a function of $V_{gate}$. (b) $dV/dI$ versus $V_{bias}$ at the Dirac point (red curve) and at high density (blue curve). (c) Current noise per unit bandwidth $S_I$ as a function of
$V_{bias}$ at the Dirac point, at $T =$ 5 K, fitted (red curve) using Khlus formula ($\mathfrak{F}$ = 0.196). (d) Mapping of the $F$ as a function of $V_{gate}$ and  $V_{bias}$ at $T =$ 5
K.}}
\end{figure}

We also measured one sample with a much smaller aspect ratio $W/L =$ 2,
having $L =$ 500 nm. The data in Fig. 4(a) shows that the minimum conductivity
in this sample is $\sim 6 \frac{e^{2}}{ h} \gg \frac{4e^{2}}{\pi h}$.
The resistance $R$ of the sample can be considered as constant as a function of the bias (see Fig. 4(b))
In Fig. 4(c), the integrated spectral density of current noise as a function
of $V_{bias}$ is fitted using Khlus formula. The fit is good at low bias,
while deviations occur at large $V_{bias}$, indicating a decrease of the Fano factor,
presumably due to electron-phonon coupling \cite{blanter2000}.
We find that the Fano factor is reduced and reaches $F =$ 0.19 at the Dirac point,
which is in agreement with the results of Ref. \onlinecite{tworzydlo2006}
calculated for the case of metallic armchair edge. $F$
as a function of $V_{bias}$ and $V_{gate}$ is displayed in Fig. 4(d).
Note that the determination of $\mathfrak{F}$ in Fig. 4(c) yields almost
the same result: $\mathfrak{F}= F = 0.19$ at the Dirac point.
We observe that $F$ is lower than expected from the theory with disorder in
Ref. \onlinecite{sanjose2007}, which strengthens the argument that transport
at the Dirac point can occur without substantial scattering.

To conclude, we have measured a gate dependent shot noise in short graphene strips with large and small $W/L$.
At the Dirac point, we observed that for large $W/L$ both minimum conductivity and Fano factor reach universal
values of $\frac{4e^2}{\pi h}$ and 1/3 respectively.
For $W/L$ smaller than 3, the Fano factor is lowered and the minimum conductivity increases.
These findings are well explained by the evanescent wave theory describing transport at the Dirac point in perfect graphene.

We thank A. Castro Neto, Y. Hancock, A. Harju, T. Heikkil\"a, A. K\"arkk\"ainen, M. Laakso, C. Lewenkopf, E. Mucciolo, M. Paalanen, P. Pasanen, and P. Virtanen for fruitful discussions. This work was supported by the Academy of Finland, the EU CARDEQ contract FP6-IST-021285-2 and the NANOSYSTEMS contract with the Nokia Research Center.


\begin{thebibliography}
\bibitem{note1} Notice that when dealing with the integrated differential Fano factor, $T_{n}$ drops out as an insignificant background offset: as a consequence, the coupling term $\frac{4k_{B}T_{n}}{Z_{0}}$ can be neglected from the Khlus formula \cite{khlus1987}.
\bibitem{note2} Note that we cannot obtain a quantitative agreement with the evanescent mode theory, probably because doping by the leads may cause variation of the gate coupling capacitance and because the presence of non-uniform doping, that, does not strongly scatter the charge carriers, also affects the electronic density of states. Nevertheless, the gate voltage scale is found to be larger than the one found in \cite{tworzydlo2006}.
\end{thebibliography}
\end{document}